\documentclass {kluwer} 

\newcommand\bb[1] {\mbox{\boldmath{$#1$}}}

\newcommand\dd{\partial}
\newcommand\beq{\begin{equation}}
\newcommand\eeq{\end{equation}}

\newcommand\bibi{ \bibitem []{} }
\newcommand\apj[3] { {\it Ap.~J.} {\bf#1}, #2--#3.}
\newcommand\apjlet[3] { {\it Ap.~J.~(Let.)} {\bf#1}, #2--#3.}
\newcommand\mn[3] { {\it M.N.R.A.S.} {\bf#1}, #2--#3.}

\begin{document}

\begin{frontmatter}

\end{frontmatter}
\begin{article}
\begin{opening}

\title{Solar Nebula Magnetohydrodynamics}

\author{Steven A. \surname{Balbus}}
\author{John F. \surname{Hawley}}
\institute {Virginia Institute of Theoretical Astronomy, Astronomy Dept., Univ.
of Virginia, Charlottesville, VA 22903--0818, USA}

\author{\footnotesize{{To be published in {\it
   ``From Dust to Terrestrial Planets,''} eds.~W. Benz, R.~Kallenbach, G.~Lugmaier, and
   F.Podosek}, ISSI Space Sciences Series No. 9, reprinted in Space Sciences Reviews (1 January 2000)}}

\institute{}

\begin{abstract}

The dynamical state of the solar nebula depends critically upon whether
or not the gas is magnetically coupled.  The presence of a subthermal
field will cause laminar flow to break down into turbulence.  Magnetic
coupling, in turn, depends upon the ionization fraction of the gas.
The inner most region of the nebula ($\lesssim 0.1$ AU) is magnetically
well-coupled, as is the outermost region ($\gtrsim 10$ AU).  The
magnetic status of intermediate scales ($\sim 1$ AU) is less certain.
It is plausible that there is a zone adjacent to the inner disk in
which turbulent heating self-consistently maintains the requisite
ionization levels.  But the region adjacent to the active outer disk is
likely to be magnetically ``dead.''  Hall currents play a significant
role in nebular magnetohydrodynamics.

Though still occasionally argued in the literature, there is simply no
evidence to support the once standard claim that differential rotation
in a Keplerian disk is prone to break down into shear turbulence by
nonlinear instabilities.  There is abundant evidence---numerical,
experimental, and analytic---in support of the stabilizing role of
Coriolis forces.  Hydrodynamical turbulence is almost certainly not a
source of enhanced turbulence in the solar nebula, or in any other
astrophysical accretion disk.

\end{abstract}


\end{opening}

\section {Introduction}

Despite decades of activity, a reliable model for the dynamical state
of the primitive solar nebula remains elusive.  It is of course
essential to have such an understanding if we are to make sense of
nebular chemistry, transport processes, planet formation and migration,
and evolutionary history.  The nebular temperature profile is of
particular importance, since much of the meteoritic data  (e.g.
condensation properties of refractory elements) bear directly upon
this.

A significant source of heat for the solar nebula will be turbulent
dissipation, at least in those regions of the disk prone to
instability.  This turbulence also results in angular momentum and
outward energy transport at rates far in excess of those associated
with molecular viscosity; one speaks therefore of an enhanced
turbulent, or ``anomalous''  viscosity.  These properties are the essence of an
accretion disk.  They allow the central star and disk to evolve, and in
some circumstances (e.g. FU Orionis outbursts) the evolution may be
eruptive.  In this contribution, we review developments in our
understanding of accretion disk turbulence as it applies to the solar
nebula, and by extension, to protostellar and protoplanetary disks more
generally.  Recent general disk reviews may be found in Papaloizou \& Lin
(1995), Lin \& Papaloizou (1996), and Balbus \& Hawley (1998).  

\section {Accretion Disk Turbulence}

\subsection {A Preliminary Comment}

For many years, it has been the custom for disk articles to identify
the source of anomalous viscosity as the principal difficulty of accretion disk
theory, use words like ``mysterious'' to describe its origin, and pay homage to 
nonlinear instabilities, convection, and possibly magnetic fields.
The author would then move on to a simple viscous model, which would form 
the basis of the discussion.

Substantial progress has been made in the last decade.  The origin of
accretion disk turbulence is not mysterious, it is in fact very
simple.  The presence of any kind of weak magnetic field (in a sense
described below) in a conducting, differentially rotating fluid renders
it violently unstable.   Remarkably, the instability was known at some
level (though its consequences were certainly not appreciated) forty
years ago (Velikhov 1959, Chandrasekhar 1960).  Its importance to
accretion disks was noted by Balbus \& Hawley (1991, 1992a,b), and its
subsequent elucidation and development have been carried through by
many investigators (e.g., Goodman and Xu 1994; Hawley, Gammie, and
Balbus 1995; Brandenburg et al. 1995; Matsumoto et al. 1996; Stone et
al. 1996; Gammie 1996; Terquem and Papaloizou 1996; Ogilvie and Pringle
1996; Balbus and Hawley 1998 for review).

If earlier work has inspired confidence in magnetohydrodynamic (MHD)
turbulence, then the most recent findings compel outright chauvinism.
Not only is it established that weakly magnetized disks become unstable,
developing enhanced turbulent transport, it also clear that hydrodynamical
turbulence is fundamentally unsuitable to serve in this role.  One may
as well try to excite convection in a top-heated fluid as excite shear
turbulence in a Keplerian disk
\footnote{This pithy observation was apparently noted by Sir Harold
Jeffreys in conversation with L. Mestel (Papaloizou, private communication).}.
Neither venture will work, ``nonlinear
instabilities'' notwithstanding.   This rather unequivocal position will
be justified later in this paper.

\subsection {The Magnetorotational Instability}

We begin with a brief presentation of the linear magnetic
(``magnetorotational'') instability, almost certainly the basis for any model
of turbulent transport in the nebula.  Set up a standard cylindrical
coordinate system $(R, \phi, z)$ centered upon a central
gravitating  body.  The equations of motion for an element of fluid
orbiting in a plane in the field of central potential $\Phi$ are
\beq\label{first}
{d^2R\over dt^2} - R \left( {d\phi\over dt}\right)^2 = - {\dd
\Phi\over \dd R} + f_R,
\eeq
\beq\label{second}
R {d^2\phi\over dt^2} + 2 {dR\over dt}{d\phi\over dt} = f_\phi,
\eeq
where $f_R$ and $f_\phi$ represent local forces in the $R$ and
$\phi$ directions.  We allow the particle to make small excursions
from a circular orbit $R=R_0$, $\phi = \Omega_0 t$ by introducing
small quasicartesian $x$ and $y$ variables
\beq\label{local}
R = R_0 + x \qquad\qquad \phi = \Omega_0 t + y/R_0.
\eeq

Substituting equations (\ref{local}) into equations (\ref{first}) and
(\ref{second}), canceling
leading order terms, and retaining those linear in $x$ and $y$
leads to the Hill equations.  These were first used to study the
Sun-Earth-Moon triplet system:
\beq
{\ddot x} - 2\Omega {\dot y} = - x {d\Omega^2\over d\ln R} + f_x,
\eeq
\beq
{\ddot y} + 2 \Omega{\dot x} = f_y.
\eeq
We follow the convention of writing dots for time derivatives and have
resolved the local force into its $x$ and $y$ components.  The angular
velocity gradient $d\Omega^2/d\ln R$ refers to the circular velocity
profile $\Omega(R)$, and we have suppressed the ``0'' subscript for
clarity.

In what follows, fluid pressure forces are unimportant, and we need
consider only magnetic stresses in the $f$-forces.  
Consider an equilibrium vertical magnetic field.
If we restrict our attention to fluid displacements
in the plane of the disk, varying only with $z$ and $t$ as $e^{i(\omega
t - k z)}$, these stresses take a very simple form:  the magnetic
tension force is $-(\bb{k\cdot u_A})^2 \bb{\xi}$ (e.g.,
Balbus and Hawley 1992a), where $\bb{u_A}$ is the Alfv\'en velocity
\beq
\bb {u_A} = { \bb{B}\over \sqrt{4\pi\rho}},
\eeq
and $\bb{\xi}$ is the two-dimensional displacement vector.  The
equations of motion take the form

\newcommand\kuA {{\bb{k\cdot u_A}}}

\beq\label{xddot}
{\ddot x} - 2\Omega{\dot y} = - x {d\Omega^2\over d\ln R} -(\kuA)^2x,
\eeq
\beq\label{yddot}
{\ddot y} + 2 \Omega{\dot x} = -(\kuA)^2y.
\eeq
These are equations for classical Alfv\'en waves (more precisely an Alfv\'en and
a slow wave, which is degenerate with the former), modified by Coriolis
forces and a radial tidal field.
The magnetic tension terms proportional to
$(\kuA)^2$ act like spring forces, with a tunable spring constant
depending upon the magnitude of $\bb{k}$.

Equations (\ref{xddot}) and (\ref{yddot}) may be satisfied provided
that $\omega$ satisfies
\beq\label{disp}
\omega^4 - \omega^2 [\kappa^2 + (\kuA)^2] +  (\kuA)^2 \left[ (\kuA)^2
+ {d\Omega^2\over d\ln R}\right] = 0,
\eeq
a simple quadratic in $\omega^2$.  Here, $\kappa^2$ is the 
epicyclic frequency,
\beq
\kappa^2 = {1\over R^3} {dR^4\Omega^2\over dR}.
\eeq
If
\beq\label{crit}
(\kuA)^2 + {d\Omega^2\over d\ln R}  <0,
\eeq
then the system will be unstable.  Physically, the left side  equation
(\ref{crit}) is (with a minus sign) the net centrifugal force on a displaced
element, the magnetic field here appearing as a restoring, stabilizing
agent.  The inequality is a statement that this force should have a radial
{\em outward\/} component for destabilization.

Assuming that large enough wavelengths
are allowable (a weak spring), when the angular {\em velocity} (not
the angular momentum) decreases outwards, the disk will be unstable.
But of course that {\it is} the behavior of a Keplerian (or any other
astrophysical disk) rotation law.  Only slightly more work is needed to
reveal that there is a maximum growth rate
\beq\label{grow}
|\omega_{max}| = - {1\over2} {d\Omega\over d\ln R}
\eeq
which, for a Keplerian rotation law $\Omega \sim r^{-3/2}$, is obtained
when
\beq\label{streng}
(\kuA)^2 = {15\over 16} \Omega^2.
\eeq
The growth rate (\ref{grow}) is quite large, amounting to an amplification
factor of about 100 per orbit.

In what sense must the field be ``weak'' for the instability to run?
Equation (\ref{crit}) provides the key.  The field must be sufficiently
weak that the Alfv\`en tension frequency is less than $\sqrt{3} \Omega$
in a Keplerian disk.  Wavenumbers in excess of this limitation are
stabilized by the tension forces.  If the field is too strong, all
wavelengths able to fit inside the thin disk will become stabilized.
Generally, this restricts the unperturbed Alfv\'en speed to be less
than of order the disk sound speed (Balbus and Hawley 1991).

To understand the instability more fully, go back to our connecting
spring analogy.  When the spring is stretched so that one mass is
orbiting slightly farther out than the other, the inner mass rotates
more rapidly on the lower orbit.  The tug of the spring pulls back on
this mass, causing it to lose angular momentum.  (This is the torque on
the right hand side of equation [\ref{yddot}]).  This loss compels the
mass to descend to a yet lower orbit, while the gain of angular
momentum in the outer mass sends it to a higher orbit.  The spring
stretches more, and the process runs away.  Note that this is an
instability only if the spring is relatively weak; if the spring is too
strong the mass points will simply oscillate as they orbit.  The
instability is strongest if the spring constant is comparable to
$\Omega^2$, as indicated by equation (\ref{streng}).  Note as well that
angular momentum transport is not some nonlinear outcome of the
instability, it is the {\em essence\/} of the instability.

Second, the instability is even more robust than our simplified
presentation indicates (Balbus and Hawley 1998).  The maximum growth
rate (\ref{grow})
is independent, not only of the strength of the magnetic field,
but also of the geometry of the magnetic field.  Even a purely toroidal
field is unstable (Balbus and Hawley 1992b; Terquem \& Papaloizou 1996),
with the same maximum growth rate (Balbus \& Hawley 1998).  The
presence of buoyancy or other pressure effects does not affect the
stability criterion, since the most unstable displacements lie in the
plane of the disk.  Furthermore, it has been conjectured that equation
(\ref{grow}) represents the maximum possible growth rate that {\it any}
instability tapping into the free energy of differential rotation can
achieve (Balbus \& Hawley 1992a).

Finally, three-dimensional numerical simulations of the nonlinear phase
of the instability (Hawley et al. 1995), combined with analytic
arguments (Balbus and Papaloizou 1999) eliminate any doubt that the
instability leads to turbulence, that the turbulence is fully
developed, and that significant angular momentum transport results.
Values of the ``$\alpha$ parameter'' (cf. next section) obtained in
local simulations extend over the range $5\times 10^{-3} \lesssim
\alpha \lesssim 5\times 10^{-1}$ (Brandenburg et al. 1995; Hawley et
al. 1995), with lower values corresponding to initially uniform toroidal
fields or fields with vanishing mean values, and the higher values
corresponding to the presence of a mean axial field.  If the weak field
MHD turbulence is present, it will act like a greatly enhanced disk
viscosity, both diffusively and dissipatively.  It will dictate the
local thermal structure of the solar nebula.  The question of
importance becomes, where in the disk will the ionization levels be
sufficiently high to permit good magnetic coupling?

\subsection {Turbulent Heating in the Nebula}

Viscous, turbulence-enhanced accretion has been
a popular and stimulating model for the post-infall global evolution of the solar nebula
(Lin and Papaloizou 1980). 
Classical viscous accretion disk models allow one to determine macroscopic disk
properties in terms of two constants: the accretion rate $\dot M$
\footnote {Note that this accretion rate refers to radial drift through the
disk, not an external infall rate.}, and a
dimensionless disk viscosity parameter, $\alpha$. 
The latter is related to the (turbulent) viscosity $\nu_T$ by
\beq
\nu_T = \alpha c_S H
\eeq
where $c_S$ is the isothermal sound speed, and $H$ is the disk scale height
(Pringle 1981).
(The quantity $c_S$ is computed with a mean mass per particle appropriate to a
molecular hydrogen gas of cosmic abundances, ${}\sim 2.33 m_p$, where $m_p$
is a proton mass.)  If, as we believe, MHD turbulence is the physics behind
$\alpha$, then it should be possible to go beyond simple parameterizations.

We know, for example, that the disk must be hot enough to maintain an ion
population capable of coupling the field to the molecular nebular gas.
Where is this requirement self-consistently met?
Using an $\alpha$ model as our starting point,
let us parameterize our results in terms of the column density
$\Sigma$ and midplane optical depth $\tau$.  The midplane nebula temperature
$T$ may be found from energy conservation:
\beq\label{disken}
{3\over4} \tau T^4 = {3GM {\dot M} \over 8 \pi R^3 \sigma},
\eeq 
and angular momentum conservation, 
\beq\label{diskmom}
{\dot M} = 3\pi \Sigma \alpha c_S H.
\eeq
(Shakura and Sunyaev 1973; Frank, King, and Raine 1992).  Here $G$ is
the gravitational constant, $M$ is the central mass, and $\sigma$ is
the Stefan-Boltzmann constant.  Equation (\ref{disken}) expresses the
condition that surface radiation losses are balanced by internal
mechanical energy dissipation; equation (\ref{diskmom}) employs a
vanishing stress boundary condition at the inner edge of the disk.

These two equations lead to a temperature of
\beq\label{temp}
T = 1140\ {\rm K}\  \left(\alpha\over 10^{-2}\right)^{1/3}
\left(\Sigma \tau\over 10^6\right)^{1/3} R_{AU}^{-1/2},
\eeq
where $R_{AU}$ is the disk radial location in astronomical units.
From the point of view of magnetic coupling,
this is an interesting temperature regime, neither prohibitively cool,
nor dominated by ions.  Clearly, different regions of the nebula will have
very different magnetic properties, very different
internal temperatures, and very different dynamical states.
The solar nebula is likely to have been highly inhomogeneous.  
Among other consequences, the ease with which a planet migrates through
the nebula depends very much upon whether the gas is turbulent or not
(Terquem, Papaloizou, and Nelson, this volume).  With the database
of new planets and even new planetary systems rapidly swelling,
understanding the extent over which protoplanetary disks maintain
turbulence is a key problem.

\section {Resistivity and Hall Currents}

Outside of the first few 0.1 AU, the ionization fraction in the solar nebula is
likely to be very small.  Of course very little ionization is needed to ensure good
coupling in astrophysical plasmas,  but even this may not always be attainable.
There are two important effects in a low ionization plasma which bear
discussion.

\subsection{Resistivity}

The resistivity of partially ionized plasma is (e.g., Blaes \& Balbus
1994)
\beq\label{eta}
\eta = 230 \left( n_n\over n_e\right) T^{1/2} \  \ {\rm cm\ s}^{-2},
\eeq
where $n_n$ is the neutral number density, $n_e$ is the electron number
density, and $T$ is the temperature.
A measure of the relative importance of the resistivity
is given by the magnetic Reynolds number
\beq\label{ReM}
Re_M = {H c_s \over \eta} \simeq {c_s^2\over \Omega \eta},
\eeq
where the final expression is our operational definition.  When $Re_M$
is smaller than $c_S^2/u_A^2$, resistivity affects the most rapidly
growing linear wavelengths, and when $Re_M$ falls below $4 \pi^2$, the
linear instability is completely quenched (Stone et al.  1999).  (Note
that $c_S^2/u_A^2$ is generally, though not inevitably, larger than
$4\pi^2$.) On scales of 1 AU in the solar nebula, this corresponds to
ionization fraction of about $10^{-13}$.  [This is a small number, but
it is large enough to be in a regime where we expect dust grains to be
playing a relatively minor role (Umebayashi \& Nakano 1988)].  The
dominant ions will be Na$^+$ and K$^+$, abundant alkalis with low
ionization potentials.

The ionization in the inner nebula will be thermal [tenuous outer
layers may be regulated by cosmic or X-rays, (Gammie 1996, Glassgold et
al. 1997)], and an ionization fraction of $10^{-13}$ is associated with
temperatures of 900--1100 K over a broad range of densities.  Taking
this result together with equation (\ref{temp}), we find that formally,
an accretion disk with $\alpha = 10^{-2}$ can maintain this minimal
ionization out to radii of order 1 AU.

We need to be careful.  Conditions for linear instability to be present
and conditions for {\em maintaining\/} turbulence need not be the
same.  (One example of this is the linear two-dimensional
magnetorotational instability, which cannot sustain turbulence because
of the anti-dynamo theorem.  Quite generally, shear turbulence requires
three dimensions.)  A local study by Fleming, Stone, \& Hawley (1999),
has shown a marked decrease in the level of MHD
turbulence for $Re_M < 5\times 10^4$, with turbulence decaying once
$Re_M< 2\times 10^4$.  These values are far too high to affect the
linear behavior of the instability, and indeed the simulations are
indistinguishable early in the linear regime.  It is possible that the
upper limit represents the discreteness of the grid, since the behavior
of $Re_M = 5\times 10^4$ and ``$Re_M =\infty$'' (i.e., no explicit
resistivity) are likewise indistinguishable.  Thus, finite magnetic
Reynolds number effects may be present at higher values as well.

$Re_M$ is extremely sensitive to the temperature.  If the ionization fraction is
much less than the abundance of K ($\sim 10^{-7}$), the Saha equation gives
\beq
{n_e\over n_n} = 6.47\times 10^{-13} a_{-7}^{1/2} T_3^{3/4} 
\left( 2.4\times 10^{15}\over n_n \right)^{1/2}
\left[\exp (-25,188/T)\over 1.15\times 10^{-11} \right]
\eeq
where $a_{-7}$ is the K abundance in unit of $10^{-7}$, $T_3$ is the
temperature in units of $10^3$ K; the Boltzmann factor is normalized to its
value at 1000 K, and the neutral density normalization is given by fundamental
constants of nature which happen to lead to a value within an order of magnitude
(or two) of that expected for the solar nebula.  
This leads to a magnetic Reynolds number of
\beq
Re_M = 15.8\, T_3^{5/4} a_{-7}^{1/2} R_{AU}^{3/2} 
\left( 2.4\times 10^{15}\over n_n \right)^{1/2}
\left[ \exp (-25,188/T)\over 1.15\times 10^{-11} \right]
\eeq
At face value this is well below what the Fleming et al. simulations suggest 
is needed for unencumbered turbulence, but the exponential Boltzmann acts like a
switch.   One need not venture more than a few tenths of an AU inward for
conditions to change dramatically, say, at $T=1400$ K.  We are then in the
regime of full MHD turbulence.  

This behavior is the most salient difference between a simple $\alpha$
model for the nebula, and one based upon MHD turbulence: a very sharp
change in the dynamical state of the inner 1 AU.
Furthermore, it is also self-consistent to have a quiescent disk as we
move inward of 1 AU, with much lower temperatures, and much lower
ionization fractions.  This would mean a smaller value for the
turbulent stress, which is the basis of the self-consistency.  More
turbulence means more ions to help maintain the magnetic coupling, less
turbulence means the converse.  Clearly, the transition between
quiescence and turbulence may be explosive: a slight increase in
temperature turns into an increase in ionization, an increase in the
alpha parameter, and yet higher temperatures.  Gammie and Menou (1998)
have argued that dwarf novae outbursts are related to just such a
transition to MHD turbulence; it is very possible that FU Orionis
outbursts are a similar phenomenon.

\subsection{The Hall Effect}

The second effect 
may be understood as follows.  In the absence of resistivity, the standard
induction equation for the magnetic field $\bb{B}$ is 
\beq
{\dd\bb{B}\over \dd t} = \bb{\nabla\times (v\times B)},
\eeq
where $\bb{v}$ is the fluid velocity.  The question is, which fluid?  
Under most circumstances, the question is moot because the difference between
ion, electron, and neutral velocities is too small to matter.  But when the
ionization fraction becomes sufficiently small (and $10^{-13}$ qualifies),
we must be careful with the distinction.  A more accurate induction equation is
\beq\label{more}
{\dd\bb{B}\over \dd t} = \bb{\nabla\times ( v_e\times B)},
\eeq
that is, the electron velocity replaces the ``fluid'' velocity.   This is
because the magnetic field lines are tied most effectively 
to the most mobile charge carriers.  In the case of interest, the inertia of
the charge carriers is negligible, and the fluid velocity is to a very good
approximation that of the neutrals.  Since
\beq\label{ve}
\bb{v_e} = \bb{v} + (\bb{v_e} - \bb{v_i}) + (\bb{v_i} - \bb{v}),
\eeq
the electron velocity may be represented as the sum of the neutral
velocity, a second term proportional to the current
($\bb{v_i}$ is the ion velocity), plus a final term
determined by the strength of ambipolar diffusion coupling.  In
interstellar molecular gas, this final term generally turns out to be
more important than the second.  But in protostellar disks, the second
term, which gives rise to the {\em Hall effect\/} in the laboratory, is
dominant.  This point is made, using the formalism of
conductivity tensors, in an important recent paper by Wardle (1999).
We will shall adopt a more dynamical approach here.

Equation (\ref{ve}) may be written
in terms of the current density $\bb{j}$, 
\beq
\bb{v_e} = \bb{v} - { \bb{j}\over n_i e},
\eeq
where $n_i$ is the ion number density (taken equal to the electron's),
and the final ion-neutral drift term is ignored.  It is clear now why
the Hall term on the right is important, with $n_i$ small and the
currents large enough to generate fields of interest.  The induction
equation (\ref{more}) takes the form
\beq
{\dd\bb{B}\over \dd t} =
\bb{\nabla\times} ( \bb{ v\times B}- { \bb{j\times B}\over n_i e}),
\eeq
which introduces Coriolis-like terms in the behavior of the field
lines.  This makes the behavior of the fluid and the field symmetric:
``field-freezing'' tries to couple the magnetic field and fluid velocities,
while Coriolis forces and Hall effect terms cause internal
epicyclic motions in the matter and field respectively.  Relative to
the disk rotation, the fluid epicyclic motion is retrograde, while the
sense of the field line motion depends on the magnetic helicity
$\bb{\Omega\cdot B}$.

A more clear picture of the effect of the Hall term emerges, if we consider
its behavior in a uniformly rotating disk, ignoring for the moment any effects
of finite electrical resistance.  If we once again consider perturbations to a
vertical field $\bb{B} = B\bb{e_z}$, and denote linear perturbations 
by $\delta$, plane wave disturbances of the form $\exp(ikz-i\omega t)$
satisfy the dynamical equations,
\beq
-i\omega \delta v_R - 2\Omega \delta v_\phi - {ikB\over 4\pi \rho}
\delta B_R = 0,
\eeq
\beq
-i\omega \delta v_\phi + 2\Omega \delta v_R - {ikB\over 4\pi \rho}
\delta B_\phi = 0,
\eeq
and the induction equations,
\beq
-i\omega\delta B_R +{k^2Bc\over 4\pi e n_i} \delta B_\phi
-ikB \delta v_R=0
\eeq
\beq
-i\omega\delta B_\phi -{k^2Bc\over 4\pi e n_i} \delta B_R
-ikb\delta v_\phi = 0.
\eeq
Note the symmetrical appearance of $\delta\bb{B}$ and $\delta
\bb{v}$ in the above, and that the Hall terms enter for the magnetic
field precisely as the Coriolis terms do for the velocity field,
except that they may induce a {\em different sense of rotation.}

The dispersion relation for the above system is
\beq\label{halldisp}
\omega^4   -\omega^2 \left[ 2k^2 u_A^2 +4\Omega^2+{k^4u_H^4\over
4\Omega^2}\right] + k^4(u_A^2 + u_H^2)^2=0
\eeq
where we have defined the Hall velocity $u_H$ by
\beq
u_H^2 = {\Omega B c\over2\pi e n_i},
\eeq
where $c$ is the speed of light.
Note that $u_H^2$ may be positive or negative depending upon the
sign of helicity factor $\bb{\Omega\cdot B}$.
The Hall term alters both the effective epicyclic frequency of the
coupled field--fluid system as well as the radial centrifugal force on
a displaced fluid element, discussed in \S 2.2.  The 
contribution of the magnetic field to the return force on a
displaced element goes from $k^2 u_A^2$
to $k^2 u_A^2 + k^2 u_H^2$, and a simple measure of the importance
of the Hall currents is the ratio $ x\equiv u_H^2/u_A^2$.

The dispersion relation in the presence of differential rotation and
resistivity is more complicated, and seems best left to the literature for
discussion (e.g., Wardle 1999).  But the stability requirements are easily stated.
In the absence of resistivity one simply replaces $u_A^2$ with $u_A^2 + u_H^2$,
\beq\label{hallstab}
k^2u_A^2(1+x) + {d\Omega^2\over d\ln R} > 0 \quad\ {\rm STABILITY}
\eeq
In the presence of a resistivity $\eta$ (units: cm sec$^{-2}$),
matters are a bit more complicated:
\beq\label{reststab}
k^2u_A^2[1+x + \zeta/(4+x)] + {d\Omega^2\over d\ln R} > 0 \quad\ {\rm STABILITY}
\eeq
where
\beq
\zeta \equiv {4\kappa^2 \eta^2\over u_A^4}.
\eeq
The effect of resistivity is to force the instability to longer wavelengths,
lessening the perturbation currents.  But once the wavelength exceeds the 
available linear dimensions, the system is stabilized.  

Overall, however, the effect of Hall currents on the stability of accretion
disks are not simple.  They both stabilize and destabilize,
depending upon whether $\bb{\Omega}$ is aligned or counter aligned with
$\bb{B}$.  One relatively straightforward effect which may be seen either
in equation (\ref{hallstab}) or (\ref{reststab}) is that when $B\rightarrow 0$,
$x \gg 1$ and the Hall term dominates.  The same magnetic tension force is
provided by much longer wavelengths than would have been possible without
the Hall currents.  This extends the range of unstable field strengths before
dissipation processes kill the instability.

We have seen that Hall effect currents affect the stability of a low
ionization protostellar disks.  How might they affect the nonlinear
development of the turbulence?  We are some ways from answering this,
but one potential complication can be addressed.  The Poynting
contribution to the turbulent energy flux is
\beq 
\bb{B\times} ( \bb{v_e}\bb {\times} \bb{B}).
\eeq
A key piece of the alpha disk formulation is that the energy flux be a
linear function of the same turbulent flow quantities responsible for
angular momentum transport:  the radial drift velocity and the $R\phi$
component of the stress tensor (Balbus \& Papaloizou 1999).  The
Poynting contribution is one of the dominant components of the energy
flux, and the fact that $\bb{v_e}$ appears above instead of $\bb{v}$
looks like a complication---we have emphasized the importance of the
difference $\bb{v} - \bb{v_e}$.  But here it is the {\em azimuthal\/}
component of the unperturbed rotational
velocity $\bb{v_e}$ which is responsible for the {\em radial\/}
contribution to the energy flux.  (What is important for a perturbation
may not be important in the unperturbed state.)
For the Hall term to be negligible in
this context, we find
\beq
u_H^2 \ll (\ell \Omega)(R \Omega)
\eeq
where $\ell$ is defined by $|\bb{\nabla \times B}| \sim B/\ell$, a sort of Taylor
microscale (e.g., Tennekes \& Lumley 1972) for MHD turbulence.  If $\ell$ 
is not much smaller than $H^2/R$, this inequality will be amply satisfied, and 
the alpha formulation of MHD turbulence should remain intact.  

\section {The Role of Self-Gravity and Hydrodynamical Processes}

\subsection{The Intermediate Disk}

Beyond 1 AU, the body of the disk is likely to be magnetically active.
$Re_M$ drops rapidly until one reaches beyond 10 AU or so,
where cosmic rays or X-rays can penetrate and leave a sufficient residual
ionization to ensure good coupling once again.  
Although the focus of this meeting has been the zone of terrestrial planets,
some mention of how the inner and outer disk are related is needed.

It is now clear that many of the usual hydrodynamical mechanisms invoked 
in the past---turbulence generated by infall, convection, etc.---will not
transport angular momentum outward.  One that certainly does work, at
least in principle, is self-gravity (e.g., Nelson et al. 1998).
If the mass of the disk within a
radius $R$ is in excess of $(H/R) M_\odot$, it will become Jeans unstable,
developing local (possibly global) spiral structure, which in turn is an
effective outward transport mechanism.  At 5 AU, the required interior disk
mass is $\sim 0.05\> M_\odot$.  This is much in excess of Jupiter's $10^{-3}\,
M_\odot$, but it is not unphysical.  It is what one might guess if planet
formation is efficient at the level of a few per cent, and consistent with the
figure of $10^{-2} \, M_\odot$ inferred for the accreted mass of the inner
disk in FU Orionis outbursts (Hartmann, Kenyon, and Hartigan 1993).  

While this behavior may be common, gravitational replenishment is not
likely to be universal, and certainly cannot go on indefinitely!  What
happens when the disk mass is too small?  In some cases, we would
expect to find passive disks with a hole on scales less than AU.   At
later stages in a disk's evolution, this will be unavoidable, as mass
is lost one way or another.  Strom, Edwards, and Skrutskie (1993) have
found evidence for disks with inner holes on precisely these scales.
They argue that these probably represent later stage transition disks
when accretion begins to cease, that the inner parts of the disk need
to be optically thin, and that the inner disk region is ``isolated''
from the outer.  Masses of the optically thick inner disks were
estimated at between 0.01 and 0.1 $M_\odot$, consistent with
self-gravity requirements.  At the time of the Strom et al. review,
planet formation was thought to be the most likely explanation.  But
among other possibilities (e.g., K\"onigl 1991), these disks may be
candidates for magnetic depletion.  An interesting prediction is that
the size of the inner hole should correlate with disk metallicity in
this picture:  higher abundances should produce bigger holes, since the
ionization levels needed for higher alphas could be maintained farther
out.  Detailed modeling is needed to test these striking general
agreements of scale and predicted morphology.

\subsection{Why Hydrodynamical Turbulence Fails}

The central feature of the picture of the Solar Nebula we have
discussed here is that turbulent transport is regulated by
magnetohydrodynamics, and that hydrodynamical transport is via
self-gravity.  If differential rotation by itself were for some reason
unstable, we would be forced to a completely different picture, one in
which the turbulence properties were much more uniform.  This point of
view retains prominent advocates (e.g., Dubrulle 1993, Richard \& Zahn
1999), and should be addressed.

The classical laboratory set-up for studying hydrodynamical turbulence
in differential rotation is to examine flow between two rotating
cylinders, the Taylor-Couette experiment.  According to the classic
text of Landau \& Lifshitz (1959, p. 110), flow predicted to be stable
by the Rayleigh criterion is shown by experiments to break down into
fully developed turbulence at sufficiently high Reynolds numbers.  The
textbook by Zel'dovich, Rumaikin, and Sokoloff (1983, p. 321) is even
more explicit on this point, arguing that the nonlinear breakdown of
{\em Keplerian\/} flow is supported by experiments.

These are extraordinary claims.  We have found no such experiments in
the literature; almost certainly none exist.  The closest approximation
to a nonlinear breakdown of Keplerian flow is seen in experiments
dominated by the rotation of the outer cylinder, or with a very small
but positive angular momentum gradient.  The first is characterized by
$d\Omega/\ln dR \gg 2 \Omega$ (shear dominating Coriolis forces), the
second by a nearly vanishing value of $\kappa^2$ (small vorticity).
Though decidedly nonlinear in their stability properties, each of these
flows differs significantly from Keplerian, which has comparable values
of $\kappa=\Omega$, $d\Omega/d\ln R= - 1.5\Omega$, and $2\Omega$.  No
Couette flow experiment has show any sign of instability when the inner
and outer cylinders follow anything close to a Keplerian profile.

To be fair, the Russian texts are somewhat misrepresentative; most
proponents of hydrodynamical turbulence would concede that the
experiments performed to date do not in fact show that Keplerian profiles are
unstable.  But, it is asserted, this is just a matter of going
to higher Reynolds number, $Re$.  Keplerian flow will be unstable at
high $Re$, because that is the way shear flows behave. 

This intuition seems born of the classical bounded planar Couette and
Poiseuille flow, and it presumes that the only dimensionless number
characterizing the flow is $Re$.  It also presumes an exquisite
sensitivity to $Re$, since proponents are forced to argue that high
resolution numerical simulations, which can easily recover the onset
of laboratory Couette instabilities, once again simply have not achieved
high enough $Re$.

To begin with, there are, of course, {\em two\/} nondimensional numbers
characterizing Couette flow, both of which influence stability.  One is
the $Re$, the other is the Rossby number $Ro$, 
\beq
Ro = \left|{1\over2}{d\ln\Omega\over d\ln R}\right|
\eeq
which measures the importance of inertial forces relative to Coriolis
forces.  Obviously, $Ro$ is a critical parameter in determining flow
stability.  As noted explicitly in the review of Bayly, Orszag, \&
Herbert (1988), the effect of viscosity on unbounded flows (this
includes shear layers and disks) is that of a {\em regular}
perturbation, and the instabilities affecting such systems are
essentially inviscid.  The viscosity, when introduced, perturbs the
flow at a level proportional to $Re^{-1} \ll 1$, not, as in boundary
layer theory, at the order unity level.  Numerical Reynolds number
effects are very small.  The Rossby number is the only inviscid flow
quantity of interest; to within a factor of 2 it is just the power law
index of $\Omega$.

The nonlinear breakdown of shear flow has been studied in some detail
(Bayly et al. 1988), both quasi-analytically and fully numerically.  In
all cases, it involves a two-stage process: an neutrally stable
nonlinear two-dimensional equilibrium is first perturbed on top of the
fundamental flow.  This is a new ``equilibrium.''   In the next stage,
this new solution is itself {\em linearly\/} perturbed in three
dimensions.  The new equilibrium is exponentially unstable.  The
existence of neutral behavior in the intermeditate two-dimension
solution is critical.  It is a feature of systems whose linear
restoring forces effectively vanish, as is the case for simple shear
layers and constant angular momentum disks.  It accounts for why
nonlinear disturbances quickly become dominant.  Systems similar to
Keplerian disks, with comparable epicyclic and shear rates, do not allow
the formation of neutrally stable nonlinear equilibria on top of the
differential rotation.

Finally, we note that turbulence maintains itself in shear layers by
vortex stretching (Tennekes and Lumley 1972).  Stretching is possible
in neutrally stable flows for which initially neighboring elements
drift apart when perturbed.  This includes disks with nearly constant
angular momentum profiles; these are indeed nonlinearly unstable.  But
in a Keplerian (or similar) disk, epicyclic motions keep elements
localized, and vortex stretching is extremely inefficient (Hawley,
Balbus, and Winters 1999).

Formal analytic arguments of the moment equations for fluctuations can
be advanced to support these arguments (Balbus \& Hawley 1998).  The
basic point is that in each of the unstable laboratory flows a dominant
source term is present in their equation, which acts obviously
differently for flows in the stable regime.  Numerical PPM simulations
on high resolution grids ($128^3$) have been performed recently (Hawley
et al. 1999) on local Keplerian flow.  The effective Reynolds number at
this resolution is very high, probably in excess of $10^5$.  The
analytic arguments were confirmed in detail, and not a trace of
instability was observed.  By way of contrast, our group has found that
nonlinear instabilities in constant angular momentum flows can be
recovered on grids as small as $8^3$!  When turbulence is directly
imposed on a Keplerian disk in the form of driven convection, the
angular momentum gradient compels {\em inward\/} angular momentum
transport (Stone \& Balbus 1996); when imposed by random dynamical
forces, no transport at all occurs (Balbus \& Hawley 1998).  It is
difficult to understand how supposedly diffusive numerics could be at
once large enough to damp Keplerian instability on a very fine grid,
small enough to allow nonlinear instabilities on an extremely crude
grid, and small enough to allow imposed thermal fluctuations to drive
{\em inward\/} angular momentum transport!  Each of these behaviors is,
however, a completely straightforward consequence of inviscid flow
dynamics.  It is simply untenable to cling to the notion that this is
all somehow a gross misinterpretation of numerical artifacts.

An angular momentum gradient in the flow is no less stabilizing than an
entropy gradient in a stratified fluid, to which it is exactly
analogous (Jeffreys 1928).  One would not argue that the radiative
interior of the sun is convectively unstable because the temperature
gradient is a source of free energy that nonlinear disturbances can tap
into.  One would not argue that the Richardson criterion for stability,
which is due to the presence of gravitational stratification across a
shear layer, is something that disappears at large Reynolds number for
nonlinear perturbations.  Angular momentum stratification stabilizes
shear flows as effectively as gravity stabilizes a bottom-heavy fluid.
A century of investigation has produced no evidence to the
contrary.

\section {An Overview of Solar Nebula Dynamics}

We are at an interesting and exciting stage in our developing understanding of
solar nebula dynamics.  Nothing like a complete model is at hand of course, but
even understanding how an isolated magnetized gas disk around a nascent sun
behaves would represent important progress.  And we {\em have\/} made progress.

We have at hand an understanding of the physical basis for generating
and maintaining turbulence, and of enhancing transport in magnetized
disks.  The magnetorotational instability has broad analytic and
numerical support.

We have not just an overall understanding of the way turbulence could
work, we also have a handle on many of the details: ionization balance,
Hall currents, nonlinear dynamics, for example.  We know that
metallicity is important, since the alkalis regulate magnetic
coupling.  There is, therefore, a direct connection between metallicity
and gas {\em dynamics\/}.  It is more usual, of course, for the metals
to regulate thermodynamical properties.

Models in which accretion, mediated by an external ionizing agent,
proceeds through the the exposed layers of the disk have been proposed
by Gammie (1996) and Glassgold et al. (1997).  However, a detailed
study of how the magnetorotational instability actually behaves for
this geometry and resistivity has not yet been done.

To summarize:  the MHD picture of the (isolated) solar nebula is one
in which turbulence is present within $\sim$ 1 AU and beyond $\sim$ 10
AU.  Gravitational instability may be present in the early 
stages on intermediate scales,
maintaining a mass influx and replenishing the inner disk.  When the
disk mass falls to the point where this is no longer possible, the
inner regions of the disk will be depleted, an effect that should manifest
itself as a deficit at near IR wavelengths.  This spectral feature 
has been observed, but an inner planet, or disruptive stellar magnetosphere
(K\"onigl 1991) are also possible interpretations.

It is a pleasure to thank C.~Terquem for several useful comments,
which lead to a significant improvement of this manuscript. 
This work has been supported by NASA grants NAG 5-7500 and NAG 5-3058.




\end{article}
\end{document}